\documentclass[12pt,a4paper,twocolumn,draft]{article}
\usepackage{cite}
\usepackage{psfig}
\newlength{\smallspace}
\newcommand{\text}[1]{\mbox{\scriptsize #1}}
\newcommand{\eq}[1]{Eq.~(\ref{eq:#1})}
\newcommand{\fig}[1]{Fig.~\ref{fig:#1}}

\newcommand{\secn}[1]{Sec.~\ref{sec:#1}}
\newcommand{\exf}[2]{\mbox{$#1\mbox{\raisebox{.1ex}{$\hspace{.7pt}\times$}} 10^{#2}$}}
\newcommand{\exff}[2]{\mbox{$#1\mbox{\footnotesize\raisebox{.2ex}{$\hspace{.4pt}\times$}} 10^{#2}$}}

\begin{document}

\renewcommand{\refname}{REFERENCES}

\thispagestyle{plain}
\begin{titlepage}

\begin{center}
 \Large\textbf{A Numerical Study of Capillary and Viscous Drainage 
in Porous Media}

\bigskip\bigskip

\normalsize
Eyvind Aker,$^{1,2,4}$ Alex Hansen$^{1,2,3}$ and Knut J\o rgen M\aa l\o y$^{4}$ 

\smallskip\smallskip

{\em $^1$ Department of Physics, University of Science and Technology,
N-7491 Trondheim, Norway}

{\em $^2$ Niels Bohr Institute, DK-2100 Copenhagen, Denmark}

{\em $^3$ Nordic Institute for Theoretical Physics, DK-2100 Copenhagen,
Denmark}

{\em $^4$ Department of Physics, University of Oslo, N-0316 Oslo, Norway}

 \end{center}
\end{titlepage}

\newpage
\setcounter{page}{1}

\section*{ABSTRACT}

This paper concentrates on the flow properties when one fluid
displaces another fluid in a two-dimensional (2D) network of pores and
throats. We consider the scale where individual pores enter the
description and we use a network model to simulate the displacement
process. 

We study the interplay between the pressure build up in the fluids and
the displacement structure in drainage. We find that our network model
properly describes the pressure buildup due to capillary and viscous
forces and that there is good correspondence between the simulated
evolution of the fluid pressures and earlier results from experiments
and simulations in slow drainage.

We investigate the burst dynamics in drain\-age going from low to high
injection rate at various fluid viscosities. The bursts are identified
as pressure drops in the pressure signal across the system. We find
that the statistical distribution of pressure drops scales according
to other systems exhibiting self-organized criticality. We compare our
results to corresponding experiments.

We also study the stabilization mechanisms of the invasion front in
horizontal 2D drainage. We focus on the process when the front
stabilizes due to the viscous forces in the liquids. We find that the
difference in capillary pressure between two different points along
the front varies almost linearly as function of length separation in
the direction of the displacement. The numerical results support new
arguments about the displacement process from those earlier suggested
for viscous stabilization. Our arguments are based on the observation
that nonwetting fluid flows in loopless strands (paths) and we
conclude that earlier suggested theories are not suitable to drainage
when nonwetting strands dominate the displacement process.  We also
show that the arguments might influence the scaling behavior of the
front width as function of the injection rate and compare some of our
results to experimental work.

\section{INTRODUCTION}

Two-phase displacements in porous media have received much attention
during the last two decades. In modern physics, the process is of
great interest due to the variety of structures obtained when changing
the fluid properties like wettability, interfacial tension,
viscosities and displacement rate. The different structures obtained
have been organized into three flow regimes: viscous
fingering~\cite{Chen-Wilk85,Maloy85}, stable
displacement~\cite{Len88}, and capillary
fingering~\cite{Len85,Cieplak88,Cieplak90}. Viscous fingering is characterized
by an unstable front of fingers that is generated when nonwetting and
less viscous fluid is displacing wetting and more viscous fluid at
relative high injection rate. The fingering structure is found to be
fractal with fractal dimension
$D=1.62$\cite{Maloy85,Chen-Wilk85}. Stable displacement is named after
the relative flat and stable front that is being generated when a nonwetting
and more viscous fluid displaces a wetting and less viscous fluid at
relative high injection rate. The last scenario, capillary fingering,
is obtained when a nonwetting fluid very slowly displaces a wetting
fluid. At sufficiently low injection rate the invasion fluid generates
a pattern similar to the cluster formed by invasion
percolation~\cite{Guyon78,Koplik82,Wilk83,Len85}. The displacement is
now solely controlled by the capillary pressure, that is the pressure
difference between the two fluids across a meniscus in a pore.

Fluid flow in porous media has also been intensively studied because
of important applications in a wide range of different technologies.
The most important areas that to a great extent depend on properties
of fluid flow in porous media, are oil recovery and hydrology.  In oil
recovery, petroleum engineers are continuously devolving improved
techniques to increase the amount of oil they are able to achieve from
the oil reservoirs. In hydrology, one important concern is often to avoid
pollution of ground water from human activity.

The simulation model presented in this paper is developed to study the
dynamics of the temporal evolution of the fluid pressures when a
nonwetting fluid displaces a wetting fluid at constant injection rate.
With the model we study the pressure in the fluids caused by the
viscous forces as well as the capillary forces due to the menisci in
the pores. The model porous medium consists of a tube network where
the tubes are connected together to form a square lattice.

Numerical simulations of fluid flow in por\-ous media using a network of
tubes was first proposed by {\em Fatt}~\cite{Fatt56} in 1956. Since then a
large number of publications related to network models and pore-scale
displacements have appeared in the
literature~\cite{Chen-Wilk85,Len88,%
Kop-Lass85,Paya86-1,King87,Blunt90,Blunt91,Reeves96,Paya96,Oeren96,Blunt97,%
Oeren97,Marck97,Blunt98,Dahle99}.  Often mentioned is the classic work
of {\em Lenormand} et al.~\cite{Len88} who were the first to
systematically classify the displacement structures into the three
flow regimes: viscous fingering, stable displacement and capillary
fingering. Including the work of {\em Lenormand} et al.,  it appears
that most network models have been used to study statistical
properties of the displacement structures or to calculate macroscopic
properties like fluid saturations and relative permeabilities.

There have been several attempts to simulate the displacement process
by using different types of growth algorithms. In 1983 {\em Wilkinson} and
{\em Willemsen}~\cite{Wilk83} formulated a new form of percolation theory,
invasion percolation (IP), that corresponds to slow drainage, i.e.\
capillary fingering. In 1984 {\em Paterson}~\cite{Pater84} was the first to
discover the remarkable parallels between diffusion-limited
aggregation (DLA)~\cite{Witten81} and viscous fingering. He also
showed similarities between anti-DLA and stable displacement. The
disadvantage with the growth algorithms is that they do not contain
any physical time and they have so far not been suitable to study the
cross over between the different flow regimes. However, attempts have
been made to use DLA and IP to study dynamics of viscous
fingering~\cite{Maloy87} and slow drainage~\cite{Maloy92,Maloy96},
respectively.

\begin{figure}
\begin{center}
\mbox{\psfig{figure=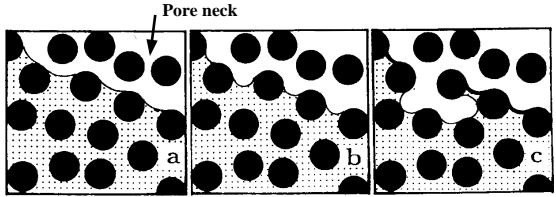,width=7.3cm}}
\caption{Nonwetting fluid (white) invades a 2D porous medium initially
filled with wetting fluid (shaded). As the nonwetting fluid is pumped
into the system (a) the menisci move into narrower parts of the pore necks
and the capillary pressure increases (b).  During a burst the invading
fluid covers new pores and the neighboring menisci readjust back to
larger radii and the capillary pressure decreases
everywhere (c)~\cite{Maloy92}. The arrow in (a) is pointing at a 
pore neck having a shape of an hourglass.}
\label{fig:burst}
\end{center}
\end{figure}

In slow drainage it is observed that the invasion of nonwetting fluid
occurs in a series of bursts accompanied by sudden negative drops in
the pressure called Haines jumps~\cite{Hain30,Maloy92,Maloy96} (see
\fig{burst}). This type of dynamics is very important for the temporal
evolution of the pressure in drainage, and in most network models
the effect is neglected. Consequently, only very few network
models~\cite{Marck97} have been used to study the interplay between
fluid pressures and displacement structures, and many questions
addressing this topic are still open.  We will attempt to answer some of
them in this paper, by making a model whose properties are closer to
those of real porous media.  To model the burst dynamics, we have been
motivated by the hourglass shaped pore necks in \fig{burst}.  As a
result we let the tubes in our network model behave as if they were
hourglass shaped with respect to the capillary pressure. Thus, the
capillary pressure of a meniscus starts at zero when the meniscus
enters the tube and increases towards a maximum value at the middle of
the tube where the tube is most narrow, before the capillary pressure
decreases to zero again when the meniscus leave the tube.

The advantage of the above approach is a network model that reproduces
the burst dynamics and the corresponding pressure evolution. We are
also able to study in detail the capillary pressure of each meniscus
along the front as it moves through the network. Similar measurements
can hardly be done experimentally.

We use the model to study the burst dynamics going from low to high
displacement rates. To do so, we examine the statistical properties of
the sudden negative pressure drops due to bursts. We find that for
a wide range of displacement rates and fluid viscosities, the pressure
drops act in analogy to theoretical predictions of systems exhibiting
self-organized criticality, such as IP. Even at high
injection rates, where the connection between the displacement process
and IP is more open, the pressure drops behave similar to the case of
extreme low injection rate, where IP apply.

Further, we report on the behavior of the capillary pressure along the
invasion front and investigate the stabilization mechanisms of
horizontal drainage. We present theoretical arguments predicting
the behavior of the pressure along the front, and we conclude that the
difference in pressure between two different points along the front
should depend almost linearly as function of the distance between the
two points in the direction of the displacement. The theoretical arguments
are based on the observation that the nonwetting fluid displaces the
wetting fluid through separate loopless strands (paths). Numerical
simulations of the capillary pressure along the front supports the
theoretical arguments. We note that earlier suggested
views~\cite{Wilk86,Len89,Blunt92,Xu98} concerning the behavior of the
pressure along the front, is not compatible with our results.

Unfortunately, the detailed modeling of the menisci's movements and their
capillary pres\-sures makes the model computationally heavy and reduces
the system size that is attainable within feasible amount of CPU time.

The paper is organized as follows. In Section~\ref{sec:model} we
present the network model and describe briefly its numerical
implementation. The rest of the sections discuss the main results that
we have got from numerical experiments with the network model. In
section~\ref{sec:press} we discuss the evolution of the simulated
pressure in drainage and calculate the statistics of the bursts. We
also compare some of our results to experimental
work. Section~\ref{sec:pcfront} presents theoretical arguments about
the stabilization mechanisms of the front in drainage.  The arguments
are supported by numerical simulations with the network model.  This
section also contains experimental work that is related to the
numerical simulations. A summary of the results and concluding remarks
are provided in Section~\ref{sec:further}.

\section{SIMULATION MODEL\label{sec:model}}
The network model is thoroughly discussed in Ref.~\cite{Aker98-1}, and
it has also been presented briefly in
Refs.~\cite{Aker98-2,Aker00-PRE}. Therefore, only its main features
are described in this section.

The model porous medium consists of a square lattice of cylindrical
tubes of length $d$ oriented at $45^\circ$ to the longest side of the
lattice. Four tubes meet at each intersection where we put a node
having no volume. The disorder is introduced by (1) assigning the
tubes a radius $r$ chosen at random inside the interval $[\lambda_1
d,\lambda_2 d]$ where $0\!\le\!\lambda_1\!<\!\lambda_2\!\le\! 1$ or
(2) moving the intersections a randomly chosen distance away from
their initial positions. The randomly chosen distances are less than
$1/2$ of the distance between the nearest neighbor intersections to
avoid overlapping nodes. In (1) all tubes have equal $d$ but
different $r$. (2) results in a distorted square lattice giving the
tubes different $d$'s. In (2) $r=d/2\alpha$ where $\alpha$ is the aspect
ratio between the tube length and its radius. The reason for making a
distorted lattice of tubes is to get closer to a real pore-throat
geometry as shown in \fig{burst}~\cite{Aker00-PRE}.

\begin{figure}
\begin{center}
\mbox{\psfig{figure=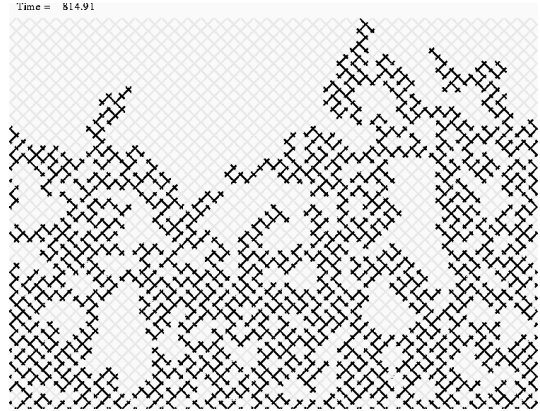,width=7cm}}
\caption{Example of a displacement structure from one simulation. The
nonwetting fluid (black) is injected from below and displaces the
wetting fluid (grey) that escapes along the top row. The front between
the nonwetting and wetting phase is defined as the line separating the
compact wetting phase and the nonwetting fluid. Note the trapped
regions of wetting fluid that are left behind and surround by
nonwetting fluid.}
\label{fig:structure}
\end{center}
\end{figure}

Figure~\ref{fig:structure} shows an example of a displacement
structure that is obtained from one simulation. The nonwetting fluid
(black) of viscosity $\mu_{nw}$ is injected along the inlet (bottom
row) and displaces the wetting fluid (grey) of viscosity
$\mu_{w}$. The fluids flow from the bottom to the top of the lattice,
and there are periodical boundary conditions in the orthogonal
direction. We assume that the fluids are immiscible and incompressible.

A meniscus is located in the tubes where nonwetting and wetting
fluids meet.  The capillary pressure $p_c$ of a meniscus in a cylindrical 
tube of radius $r$ is given by Young-Laplace law,
\begin{equation}
p_c=\frac{2\gamma}{r}\cos\theta ,
\label{eq:pc-1}
\end{equation}
under the assumption that the principal radii of the
curvature of the meniscus are equal to the radius of the tube. $\theta$ denotes
the wetting angle between the cylinder wall and the wetting fluid, i.e.
$0^\circ\le\theta < 90^\circ$ in drainage. 

In the network model we treat the tubes as if they were hourglass shaped with
respect to the capillary pressure. Therefore, we let the capillary pressure 
depend on where the meniscus is situated in the tube. Instead
of \eq{pc-1} we let $p_c$ of a meniscus vary in the following way:
\begin{equation}
p_c=\frac{2\gamma}{r}\left[ 1-\cos \left(2\pi\mbox{\large $\frac{x}{d}$}\right)\right].
\label{eq:pc-2}
\end{equation}
Here we have assumed that the wetting fluid perfectly wets the medium, i.e.\
$\theta=0$. In the above relation $x$ denotes the position of the
meniscus in the tube ($0\le x \le d$), giving that $p_c=0$ at the
entrance and at the exit of the tube and reaches a maximum of $4\gamma/r$
in the middle of the tube ($x=d/2$). Practically, the wetting angle of
a meniscus and thereby its capillary pressure may generally be
different depending on whether the meniscus retires from or invades
the tube. To avoid numerical complications this effect is neglected in
the present model.

We solve the volume flux through each tube by using Hagen-Poiseuille
flow for cy\-lindrical tubes and the Washburn approximation~\cite{Wash21}
for menisci under motion. Let $q_{ij}$ denote the volume flux through
the tube from the $i$th to the $j$th node, then we have
\begin{equation}
q_{ij}=-\frac{\sigma_{ij} k_{ij}}{\mu_{ij}}\frac{1}{d_{ij}}(\Delta p_{ij}-p_{c,ij}).
\label{eq:tubeflow}
\end{equation}
Here $k_{ij}$ is the permeability of the tube ($r_{ij}^2/8$) and
$\sigma_{ij}$ is the cross section ($\pi r_{ij}^2$) of the tube. $\mu_{ij}$
denotes the effective viscosity, that is the sum of the volume
fractions of each fluid inside the tube multiplied by their respective
viscosities. The pressure drop across the tube is $\Delta p_{ij}=p_j-p_i$,
where $p_i$ and $p_j$ is the pressures at node $i$ and $j$, respectively.
The capillary pressure $p_{c,ij}$ is the sum of the capillary pressures 
of each menisci [given by \eq{pc-2}] that are present inside the tube.
A tube partially filled with both liquids is allowed to contain at maximum 
two menisci. For a tube without menisci, $p_{c,ij}=0$. We only consider
horizontal flow, and therefore we neglect gravity.

We have conservation of volume flux at each node giving
\begin{equation}
\sum_j q_{ij}=0.
\label{eq:Kirch}
\end{equation}
The summation on $j$ runs over the nearest neighbor 
nodes to the $i$th node while $i$ runs over all nodes that do not 
belong to the top or bottom rows, that is, the internal nodes. 
Eqs.~(\ref{eq:tubeflow}) and~(\ref{eq:Kirch}) constitute a set of
linear equations which we solve for the nodal pressures $p_i$, with
the constraint that the pressures at the nodes belonging to the upper
and lower rows are kept fixed.  The set of equations is solved by
using the Conjugate Gradient method~\cite{Bat88}.

In the simulations we impose the injection rate $Q$ on the inlet,
therefore we have to find the pressure across the lattice $\Delta P$,
that corresponds to the given $Q$. Having found $\Delta P$ we use this
pressure to calculate the correct $p_i$'s in \eq{tubeflow}.  In short,
we find $\Delta P$ by considering the relation
\begin{equation}
Q=A\Delta P+B.
\label{eq:Darcy}
\end{equation}
The first part of \eq{Darcy} results from Darcy's law for single phase
flow through porous media. The second part comes from the capillary
pressure between the two fluids (i.e. $B=0$ if no menisci are present
in the network). \eq{Darcy} has two unknowns, $A$ and $B$, which we
calculate by solving \eq{Kirch} twice for two different applied
pressures $\Delta P'$ and $\Delta P''$, across the lattice.  From
those two solutions we find the corresponding injection rates $Q'$ and
$Q''$.  Inserting $Q'$, $Q''$, $\Delta P'$, and $\Delta P''$ into
\eq{Darcy} results in two equations which we solve for $A$ and
$B$. Finally, we find the correct $\Delta P$ due to the imposed $Q$ by
rewriting \eq{Darcy}, giving $\Delta P=(Q-B)/A$. See
Refs.~\cite{Aker98-1,Aker98-2} for further details on how the $p_i$'s are
calculated after $\Delta P$ is found.

Given the correct solution of $p_i$ we calculate the volume flux
$q_{ij}$ through each tube in the lattice, using \eq{tubeflow}. Having
found the $q_{ij}$'s, we define a time step $\Delta t$ such that every
meniscus is allowed to travel at most a maximum step length $\Delta
x_{\mbox{\scriptsize max}}$ during that time step. Each meniscus is
moved a distance $(q_{ij}/\sigma_{ij})\Delta t$ and the total time
lapse is recorded before the nodal pressures $p_i$, are solved for the
new fluid configuration. Menisci that are moved out of a tube during a
time step are spread into neighboring tubes as described in 
Refs.~\cite{Aker98-1,Aker98-2}.

\section{TEMPORAL EVOLUTION OF FLUID PRESSURE\label{sec:press}}

To characterize the different fluid properties used in the
simulations, we use the capillary number $C_a$ and the viscosity ratio
$M$. The capillary number indicates the ratio between viscous and
capillary forces and in the simulations it is defined as
\begin{equation}
C_a\equiv\frac{Q\mu}{\Sigma\gamma}.
\label{eq:Ca}
\end{equation}
Here $Q$ is the injection rate of the nonwetting fluid, $\mu$ is the
maximum viscosity of the nonwetting and wetting fluid and $\Sigma$ is
equal to the length of the inlet times the average thickness of the
lattice, i.e.\ $\Sigma$ is the cross section of the inlet. $\gamma$ is
the fluid-fluid interface tension.

The viscosity ratio $M$, is defined as
\begin{equation}
M\equiv\frac{\mu_{nw}}{\mu_{w}},
\label{eq:M}
\end{equation}
where $\mu_{nw}$ and $\mu_w$ is the viscosity of the invading
nonwetting fluid and the defending wetting fluid, respectively.

\subsection{TRAPPED FLUID AND PRESSURE BUILDUP\label{sec:interplay}}

The pressure across the system is found from \eq{Darcy} giving
\begin{equation}
\Delta P = \frac{Q}{A} + P_{cg},
\label{eq:press}
\end{equation}
where $P_{cg}\equiv -B/A$ defines the global capillary pressure of the
system.  As will become clear below, $P_{cg}$ contains the capillary
pressures of the menisci surrounding the trapped wetting fluid (cluster
menisci) and the capillary pressures of the menisci along the invasion
front (front menisci) (see Fig.~\ref{fig:structure}).

\begin{figure}
\begin{center}
\mbox{\psfig{figure=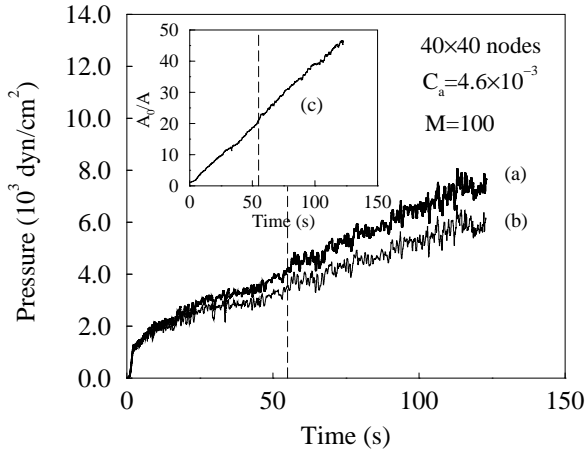,width=7.5cm}}
\caption{$\Delta P$ (a), $P_{cg}$ (b), and $A_0/A$ (c) as function of
injection time. $C_a=\protect\exff{4.6}{-3}$ and
$M=100$. The vertical dashed line is drawn at the
saturation time, $t_s$.}
\label{fig:press_sd}
\end{center}
\end{figure}
Figure~\ref{fig:press_sd} shows the simulated pressures $\Delta P$ and
$P_{cg}$ for a displacement at $C_a=\exf{4.6}{-3}$ and $M=100$.  The
front width was observed to stabilize after some time $t_s$, and a
typical compact pattern of small clusters of wetting fluid developed
behind the front.  From \fig{press_sd} we observe that both $\Delta P$
and $P_{cg}$ increases as the more viscous fluid is pumped into the
system. When $t>t_s$ they even tend to increase linearly as function
of time.

The driving mechanism in the displacement is the pressure gradient
between the inlet and the front causing a viscous drag on the trapped
clusters. At moderate injection rates these clusters are immobile,
thus the viscous drag is balanced by capillary forces along the
interface of the cluster.  On average the sum of the capillary forces
from each cluster contributes to $P_{cg}$ by a certain amount making
$P_{cg}$ proportional to the number of clusters behind the
front. After the front has saturated with fully developed clusters
behind ($t>t_s$), the number of clusters are expected to increase
linearly with the amount of injected fluid. Since the injection rate
is held fixed we recognize that $P_{cg}$ must increase linearly as
function of time. The argument does not apply when $t<t_s$, due to the
fractal development of the front before saturation.

In \fig{press_sd} we have also plotted $A_0/A$ which is the normalized
difference between $\Delta P$ and $P_{cg}$ [see \eq{press}].  $A_0$ is
equal to the proportionality factor between $Q$ and $\Delta P$ when
only one phase flows through the lattice (i.e. $P_{cg}=0$). We observe
that $A_0/A$ tends to increase linearly as function of time when
$t>t_s$. From \eq{Darcy} we interpret $A$ as the total conductance of
the lattice, and the reciprocal of that is the total resistance.  The
total resistance depends on the fluid configuration and the geometry
of the network. Locally, the fluid configuration changes as nonwetting
fluid invades the system, however, the linear behavior of $A_0/A$
indicates that the overall displacement structure is statistically
invariant with respect to the injection time. That means, after the
front has saturated ($t>t_s$) the displacement structure might be
assigned a constant resistance per unit length.

In the special case when $M=1$ (viscosity matched fluids) the total
resistance, $1/A$, was found to be constant independent of the
injection rate or displacement structure. This somewhat surprising
result might be explained by the following consideration. When
$M=1$ the effective viscosity $\mu_{ij}$, of each tube is independent
of the amount of wetting and nonwetting fluid that occupies the
tube. Hence, each tube has a constant mobility of
$k_{ij}/\mu_{ij}$ giving a constant total resistance of the network.

\begin{figure}
\begin{center}
\mbox{\psfig{figure=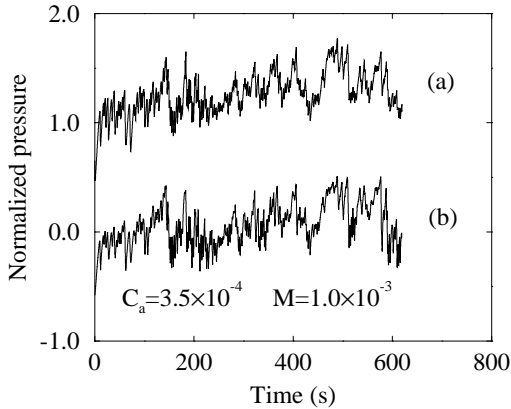,width=7cm}}
\caption{$P_{c\hspace{-.5pt}f}$ (a) and $P_{cg}$ (b) as function of
injection time at $C_a=\protect\exff{3.5}{-4}$ and
$M=\protect\exff{1.0}{-3}$. To avoid overlapping curves $P_{cg}$ was
subtracted by $1000\,\mbox{dyn}/\mbox{cm}^2$ before it was normalized.}
\label{fig:press_cf}
\end{center}
\end{figure}

At low $C_a$ we approach the regime of capillary fingering and the
viscous drag on the clusters becomes negligible. Hence, $P_{cg}$ is no
longer a linear function of the injection time, but reduces to that
describing the capillary pressure along the front. This is observed in
\fig{press_cf} where we compare $P_{cg}$ with the calculated average
capillary pressure along the front, $P_{c\hspace{-.5pt}f}$. In the
simulations, $P_{c\hspace{-.5pt}f}$ is calculated by taking the mean
of the capillary pressures of the front menisci. From the figure we
see that $P_{cg}\simeq P_{c\hspace{-.5pt}f}$, as expected. The big
jumps in the pressure functions in \fig{press_cf} are caused by the
variations in the capillary pressure as menisci move through the
``hourglass'' shaped tubes. The negative jumps are identified as
bursts where a meniscus proceeds abruptly~\cite{Hain30,Maloy92} to
fill the tube with nonwetting fluid (see also \secn{burst} for further
details).

From the above discussion we conclude that the behavior of $P_{cg}$
at large times ($t>t_s$) may be formulated as
\begin{equation}
P_{cg}=\Delta_{mc}h+P_{m\hspace{-.5pt}f},
\label{eq:Pcg}
\end{equation}
where $\Delta_{mc}$ is the proportionality factor between $P_{cg}$ and
the average front position $h$ above the inlet. $\Delta_{mc}$ is given
by the viscous drag on the clusters and $P_{m\hspace{-.5pt}f}$ is
the variation in the capillary pressure when the invasion front covers
new tubes.  $h$ is only defined after the front has saturated, i.e.
$h_s < h < L$, where $h_s$ is the average front position at $t_s$ and
$L$ is the length of the system. Since the injection rate is held
fixed, $h$ is proportional to the injection time $t$. In the limit of
very low injection rates, $\Delta_{mc}\rightarrow 0$.

When the average front position has reach\-ed the outlet, i.e. $h=L$ in
\eq{Pcg}, only invading fluid flows through the system and
$P_{m\hspace{-.5pt}f}=0$. In this limit Darcy's law applied on the
nonwetting phase gives $U=(K_e/\mu_{nw})(\Delta P/L)$, where $K_e$ is
the effective permeability of the nonwetting phase.  From
Eqs.~(\ref{eq:press}) and~(\ref{eq:Pcg}) we find that $\Delta
P=Q/A+\Delta_{mc}L$, which inserted into the Darcy equation gives
\begin{equation}
K_e=\frac{\mu_{nw}}{1/\sigma_T + \Delta_{mc}/U}.
\label{eq:Ke}
\end{equation}
Here $\sigma_T\equiv AL/\Sigma$ denotes the total conductivity of the
lattice. Thus, we might consider the effective permeability of the
nonwetting phase as a function of the conductivity of the lattice and
an additional term due to the viscous drag on the clusters
($\Delta_{mc}/U$). Note that the $U$ dependency in \eq{Ke} only
indicates changes in $\Delta_{mc}$ between displacements executed at
different injection rates. The behavior when the flow rate changes
during a given displacement is not discussed here.

\subsection{BURST DYNAMICS\label{sec:burst}}

\begin{figure}
\begin{center}
\mbox{\psfig{figure=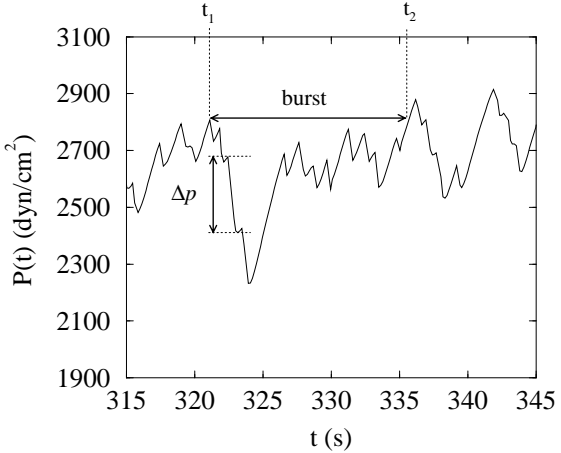,width=7.3cm}}
\caption{The pressure signal as function of injection time, $P(t)$,
for one simulation at low displacement rate in a narrow time
interval. The horizontal line defines the pressure valley of a burst
that last a time period $T=t_2-t_1$. Note that the valley may contain
a hierarchical structure of smaller valleys inside. The vertical line
indicates the size of a local pressure jump $\Delta p$ inside the
valley.}
\label{fig:burst_def}
\end{center}
\end{figure}

In the simulations a burst starts when the pressure drops suddenly
and stops when the pressure has raised to a value above the pressure
that initiated the burst (see \fig{burst_def}). Thus, a burst may
consist of a large pressure valley containing a hierarchical structure
of smaller pressure jumps ({\it i.e.}\ bursts) inside.
A pressure jump, indicated as $\Delta p$ in \fig{burst_def}, is the
pressure difference from the point when the pressure starts
decreasing minus the pressure when it stops decreasing. We define the
size of the pressure valley (valley size) to be
$\chi\!\equiv\sum_i\Delta p_i$, where the summation index $i$ runs
over all the pressure jumps $\Delta p_i$ inside the valley. The
definition is motivated by experimental work in Ref.~\cite{Maloy96}. For
slow displacements we have that $\chi$ is proportional to the
geometric burst size $s$, being invaded during the pressure
valley. This statement has been justified in Ref.~\cite{Maloy96},
where it was observed that in stable periods, the pressure increased
linearly as function of the volume being injected into the
system. Later, in an unstable period where the pressure drops abruptly
due to a burst, this pressure drop is proportional to $s$. At fast
displacements the pressure may no longer be a linear function of the
volume injected into the system. Therefore, a better estimate of $s$
there, is to compute the time period $T$ of the pressure valley
(\fig{burst_def}).  Since the displacements are performed with
constant rate, it is reasonable to assume that $T$ is always
proportional to the volume being injected during the valley and hence,
$T\propto s$.

\begin{figure}
\begin{center}
\vspace{.4cm}\mbox{}
\mbox{\psfig{figure=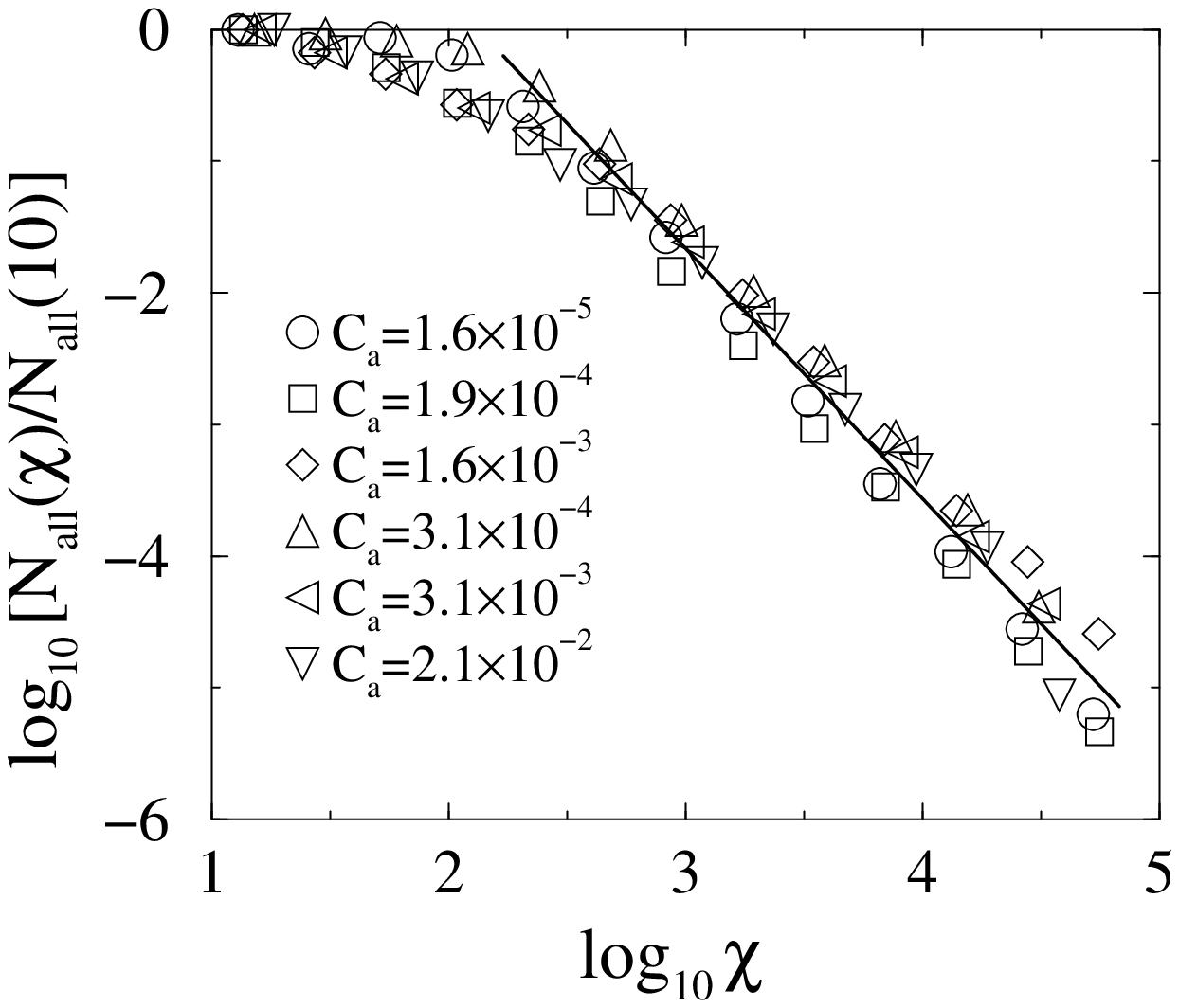,width=6.8cm}}
\caption{The hierarchical valley size distributions
$N_{\rm all}(\chi)$, for six simulations between low and high $C_a$
with $M=1$
(\mbox{\Large\protect\raisebox{-.1ex}{$\circ$}}\,,\,\put(1,0){\protect\framebox(5.1,5.1){}}\
\,\,\,,\mbox{\,\Large\protect\raisebox{-.1ex}{$\diamond$}}) and
$M=100$
($\bigtriangleup\,$,\mbox{\,\Large$\triangleleft\,$},\mbox{\,\protect\raisebox{.5ex}{$\bigtriangledown$}}). The
slope of the solid line is $-1.9$.} 
\label{fig:hier_burst}
\end{center}
\end{figure}
In \fig{hier_burst} we have plotted the hierarchical valley size
distribution $N_{\rm all}(\chi)$, for six simulations between low
and high $C_a$ with $M=1$ and $100$ on a lattice of $40\times 60$ and
$25\times 35$ nodes, respectively. $N_{\rm all}(\chi)$ was
calculated by including all valley sizes and the hierarchical smaller
ones within a large valley (see Fig.~\ref{fig:burst_def}). To obtain
reliable average quantities we did 10 to 20 different simulations at
each $C_a$. In order to calculate the valley sizes at large $C_a$, we
subtract the average drift in the pressure signal due to viscous
forces such that the pressure becomes a function that fluctuates
around some mean pressure.

By assuming a power law $N_{\rm all}(\chi)\propto\chi^{-\tau_{\rm
all}}$ our best estimate from \fig{hier_burst} is $\tau_{\rm all}=1.9\pm
0.1$, indicated by the slope of the solid line. At low $\chi$ in
\fig{hier_burst}, typically only one tube is invaded during the valley
and we do not expect the power law to be valid. Similar results were
obtained when calculating the hierarchical distribution of the time
periods $T$ of the valleys, denoted as $N_{\rm all}(T)$.

In invasion percolation (IP) the distribution of burst sizes $N(s)$,
where $s$ denotes the burst size, is found to obey the scaling
relation~\cite{Maloy92,Maloy96,Sap89,Cieplak91-1}
\begin{equation}
N(s)\propto s^{-\tau'}g(s^\sigma (f_0-f_c)).
\label{eq:burst_dist}
\end{equation}
Here $f_c$ is the percolation threshold of the system and $g(x)$ is
some scaling function, which decays exponentially when $x\gg 1$ and is
a constant when $x\rightarrow 0$. $\tau'$ is related to percolation
exponents like $\tau'=1+D_f/D-1/(D\nu)$~\cite{Cieplak91-1}, where
$D_f$ and $D$ is the fractal dimension of the front and the mass of
the percolation cluster, respectively. $\nu$ is the correlation length
exponent in percolation theory and $\sigma=1/(\nu
D)$~\cite{Stauf92}. In \eq{burst_dist} a burst is defined as the
connected structure of sites that is invaded following one root site
of random number $f_0$, along the invasion front. All sites in the
burst have random numbers smaller than $f_0$, and the burst stops when
the random number of the next site to be
invaded is larger than $f_0$~\cite{Roux89}.

By integrating \eq{burst_dist} over all $f_0$ in the interval
$[0,f_c]$ {\em Maslov}~\cite{Maslov95} deduced a scaling relation for the
hierarchical burst size distribution $N_{\rm all}(s)$ following
\begin{equation}
N_{\rm all}(s)\propto s^{-\tau_{\rm all}},
\label{eq:burst_hier}
\end{equation}
where $\tau_{\rm all}=2$.  

In the low $C_a$ regime in \fig{hier_burst}, the displacements are in
the capillary dominated regime and the invading fluid generates a
growing cluster similar to IP~\cite{Wilk83,Len85,Guyon78,Koplik82}. In
this regime we also have that $\chi\propto s$~\cite{Maloy96} and hence
$N_{\rm all}(\chi)$ corresponds to $N_{\rm all}(s)$ in
\eq{burst_hier}. Thus, in the low $C_a$ regime we expect that $N_{\rm
all}(\chi)$ follows a power law with exponent $\tau_{\rm all}=2$ which
is confirmed by our numerical results. Similar results were obtained
in Ref.~\cite{Maloy96}.

The evidence in \fig{hier_burst}, that $\tau_{\rm all}$ does not seem
to depend on $C_a$, is very interesting. At high $C_a$ when
$M=0.01$ an unstable viscous fingering structure generates and when
$M\ge1$ a stable front develops.  It is an open question how these
displacement processes map to the proposed scaling in
\eq{burst_hier}. We note that in the high $C_a$ regime the relation
$\chi\propto s$ may not be correct and $T$ is preferred when computing
$N_{\rm all}$. However, the simulations show that $N_{\rm
all}(\chi)\sim N_{\rm all}(T)$ even at high $C_a$.

In Ref.~\cite{Maslov95} $\tau_{\mbox{\scriptsize all}}$ was pointed
out to be super universal for a broad class of self-organized critical
models including IP. The result in \fig{hier_burst} indicates that the
simulated displacements might belong to the same super universality
class even at high injection rates where there is no clear mapping
between the displacement process and IP.

{\em Basak} et al.~\cite{Basak96,Aker00-Euro} performed four drain\-age
experiments where they used a $110\times 180$ mm transparent porous
model consisting of a mono-layer of randomly placed glass beads of 1
mm, sandwiched between two Plexiglas plates. The experimental setup was
similar to the one used in Ref.~\cite{Maloy92}. The model was
initially filled with a water-glycerol mixture of viscosity 0.17
P. The water-glycerol mixture was withdrawn from one of the short side
of the system at constant rate by letting air enter the system from
the other short side.  The pressure in the water-glycerol mixture on
the withdrawn side was measured with a pressure sensor of our own
construction.

From the recorded pressure signal we calculated the hierarchical
distribution of time periods of the valleys, $N_{\rm all}(T)$. At low
$C_a$ this corresponds to $N_{\rm all}(s)$ in \eq{burst_hier}.
Because of the relative long response time of the pressure sensor,
rapid and small pressure jumps due to small bursts are presumably
smeared out by the sensor and the recorded pressure jumps are only
reliable for larger bursts. Hence, from the recorded pressure signal
$T$ appears to be a better estimate of the burst sizes than $\chi$.

\begin{figure}
\begin{center}
\mbox{\psfig{figure=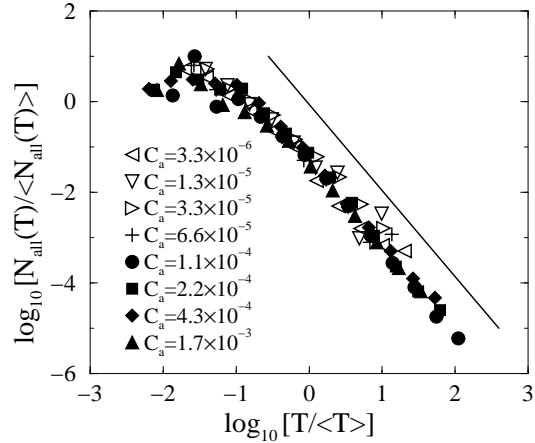,width=7cm}}
\caption{The hierarchical distribution $N_{\rm
all}(T)$ of the valley time $T$ during a burst for experiments (open
symbols) and simulations (filled symbols) at various $C_{\rm a}$ with
$M=0.017$ and $M=0.01$, respectively. The slope of the solid line is
$-1.9$.}
\label{fig:hier_top}
\end{center}
\end{figure}
In \fig{hier_top} we have plotted the logarithm of $N_{\rm all}(T)$
for experiments (open symbols) and simulations (filled symbols)
performed at four different $C_{\rm a}$, respectively. To collapse the 
data $N_{\rm all}(T)$ and $T$ were normalized by their means. In the
simulations $M=0.01$ while in the experiments $M=0.017$ where we have
assumed air to have viscosity \exf{0.29}{-2} P. We observe that the
experimental result is consistent with our simulations and we conclude
that $N_{\rm all}(T)\propto T^{1.9\pm 0.1}$. This confirms the scaling
of $N_{\rm all}(\chi)$ in \fig{hier_burst}. 

From Figs.~\ref{fig:hier_burst} and~\ref{fig:hier_top} we conclude
that $\tau_{\rm all}=1.9\pm 0.1$ for all displacement simulations
going from low to high injection rates when $M=0.01$, $1$, and
$100$. This is also confirmed by drainage experiments performed at
various injection rates with $M=0.017$.  The evidence that $\tau_{\rm
all}$ is independent of the injection rate, may indicate that the
displacement process belongs to the same super universality class as
the self-organized critical models in Ref.~\cite{Maslov95} ($\tau_{\rm
all}=2$), even where there is no clear mapping between the displacement
process and IP.

\section{STABILIZATION MECHANISMS OF THE FRONT\label{sec:pcfront}}

When the displacements are oriented out of the horizontal plane and
the injection rate is low, gravity acting on the system, may stabilize
the front due to density differences between the fluids. Several
authors~\cite{Wilk84,Wilk86,Birov91,Birov92} have confirmed, by
experiments and simulations, that the saturated front width $w_s$
scales with the strength of gravity like $w_s\propto
{B_o}^{-\nu/(1+\nu)}$. Here $B_o$ (Bond number) is the ratio between
gravitational and capillary forces, given by $B_o=\Delta\rho g
a^2/\gamma$, where $\Delta\rho$ is the density difference between the
fluids, $g$ the acceleration due to gravity, $a$ the average pore
size, and $\gamma$ the fluid-fluid interface tension. Furthermore,
$\nu$ denotes the correlation length exponent in percolation.

A similar consensus concerning the stabilization mechanisms when the
displacements are oriented within the horizontal plane, has not yet been
reached.  In horizontal displacements, viscous forces replace
gravitational forces, and in the literature there exist different
suggestions about the scaling of $w_s$ as function of $C_a$. The
capillary number $C_a$ is the ratio between viscous and capillary
forces according to the definition in \eq{Ca}. In 3D, where
trapping of wetting fluid is assumed to be of little importance,
{\em Wilkinson}~\cite{Wilk86} was the first to use percolation to deduce a
power law like $w_s\propto {C_a}^{-\alpha}$ where
$\alpha=\nu/(1+t-\beta+\nu)$.  Here $t$ and $\beta$ is the
conductivity and order parameter exponent in percolation,
respectively. Later, {\em Blunt} et al.~\cite{Blunt92} suggested in 3D
that $\alpha=\nu/(1+t+\nu)$. This is identical to the result of
{\em Lenormand}~\cite{Len89} finding a power law as function of system size
for the domain boundary in the $C_a$--$M$ plane between capillary
fingering and stable displacement in 2D porous media.

\begin{figure}
\begin{center}
\mbox{\psfig{figure=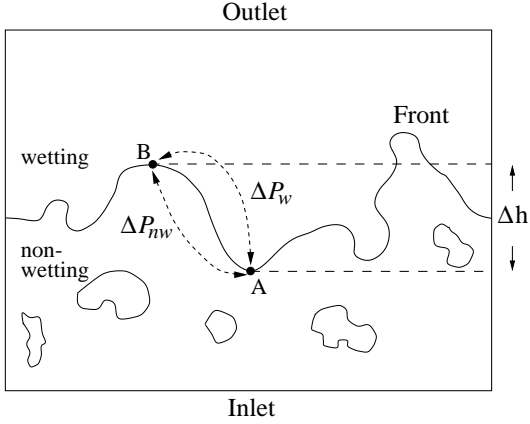,width=7cm}}
\caption{A schematic picture of the front that travels across the
system from the inlet to the outlet.  $\Delta P_{nw}$ and $\Delta P_w$
denote the viscous pressure drop going from $A$ to $B$ in the
nonwetting and wetting phase, respectively. $A$ and $B$ are separated
a distance $\Delta h$ along the short side of the system.}
\label{fig:model}
\end{center}
\end{figure}

More recently, {\em Xu} et al.~\cite{Xu98} used the arguments of
{\em Gouyet} et al.~\cite{Sap88} and {\em Wilkinson}~\cite{Wilk86} to show
that the pressure drop $\Delta P_{nw}$ over a length $\Delta h$ in the
nonwetting phase of the front should scale as $\Delta P_{nw}\propto
\Delta h^{t/\nu+D_{\mbox{\tiny E}}-1-\beta/\nu}$ (see
\fig{model}). Here $D_{\text{E}}$ denotes the Euclidean dimension of
the space in which the front is embedded, i.e.  in our case
$D_{\text{E}}=2$. The pressure drop in the wetting phase $\Delta
P_{w}$, was argued to be linearly dependent on $\Delta h$ due to the
compact phase there (\fig{model}). In Ref.~\cite{Blunt92} {\em Blunt}
et al. also suggested a scaling relation for $\Delta P_{nw}$,
however, in 3D they found $\Delta P_{nw}\propto\Delta
h^{t/\nu+1}$. This is different from the result of {\em Xu} et al.
when $D_{\text{E}}=3$.

In the next section, \secn{loopless}, we present an alternative view
on the displacement process from those initiated by
{\em Wilkinson}~\cite{Wilk86}, but include the evidence that nonwetting
fluid flows in separate strands (see \fig{strands}). The alternative
view leads to another scaling of $\Delta P_{nw}$ than the one {\em Xu}
et al.~\cite{Xu98} would predict in 2D, and we show that it may
influence $\alpha$ in the scaling between $w_s$ and $C_a$. The new
scaling of $\Delta P_{nw}$ is supported by numerical experiments using
the network model.

\subsection{LOOPLESS STRANDS\label{sec:loopless}}
\begin{figure}
\begin{center}
\mbox{\psfig{figure=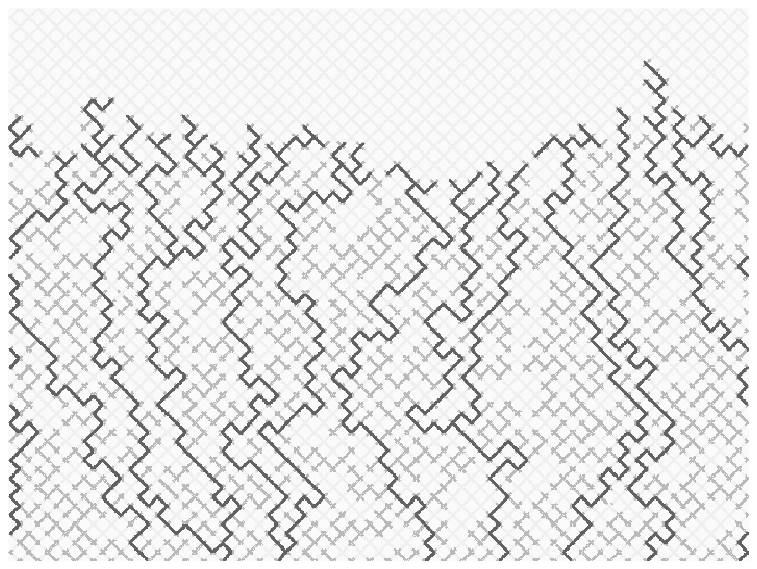,width=7cm}}\vspace{.6cm}

\mbox{\psfig{figure=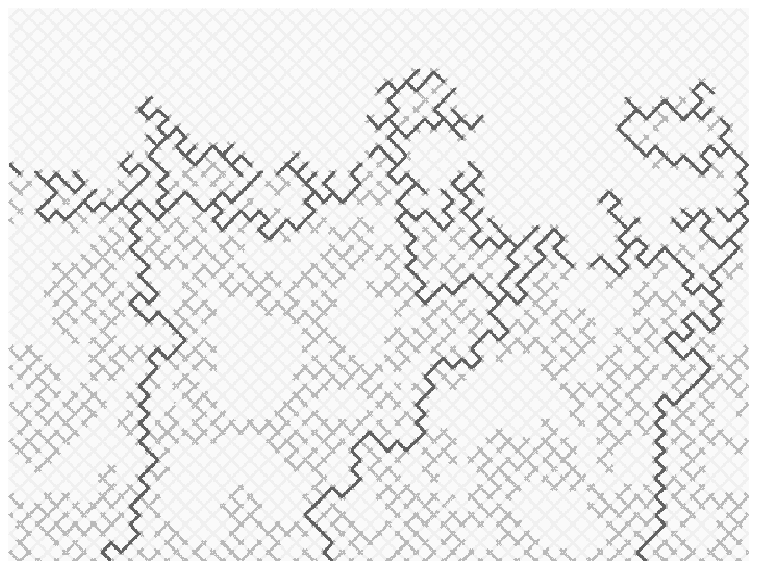,width=7cm}}
\caption{Two displacement structures of simulations at high
$C_a=\protect\exff{3.9}{-4}$ (top) and low $C_a=\protect\exff{1.6}{-5}$
(bottom) before breakthrough of nonwetting fluid. The lattice size is
$40\times 60$ nodes and $M=1$. The nonwetting fluid (dark grey and
black) is injected from below and wetting fluid (light grey) flows out
along the top row.  The black tubes denote the loopless strands where
nonwetting fluid flows and the dark grey tubes indicate nonwetting
fluid unable to flow (i.e. dead ends) due to trapped regions of
wetting fluid. Note the few fluid supplying strands from the inlet to
the frontal region at low $C_a$ compared to the case at high $C_a$.}
\label{fig:strands}
\end{center}
\end{figure}
Figure~\ref{fig:strands} shows two typical displacement structures
that were obtained from simulations at low and high $C_a$ on a lattice
of $40\times 60$ nodes with $M=1$. From the figure we observe that the
nonwetting fluid (dark grey and black) generates patterns containing
no closed loops. That means, following a path of nonwetting fluid will
never bring us back to the starting point. The nonwetting fluid also
flows in separate loopless strands, indicated as black tubes in
\fig{strands}.  The loopless structures in \fig{strands} are a direct
consequence of the evidence that a tube filled with wetting fluid and
surrounded on both sides by nonwetting fluid is trapped due to volume
conservation of wetting fluid~\cite{Aker00-PRL}. We note that this
evidence may easily be generalized to 3D, and therefore our arguments
should apply there too.  We also note that trapping of wetting fluid
is more difficult in real porous media due to a more complex topology
of pores and throats there. Loopless IP patterns have earlier been
observed in Refs.~\cite{Sahimi98,Yortsos97,Cieplak96}.

From \fig{strands} we may separate the displacement patterns into two
parts. One consisting of the frontal region continuously covering new
tubes, and the other consisting of the more static structure behind
the front. The frontal region is supplied by nonwetting fluid through
a set of strands that connect the frontal region to the inlet. When
the strands approach the frontal region they are more likely to
split. Since we are dealing with a square lattice, a splitting strand
may create either two or three new strands.  As the strands proceed
upwards in \fig{strands}, they split repeatedly until the frontal region
is completely covered by nonwetting strands.

On IP patterns without loops~\cite{Sahimi98,Porto97,Cieplak96} the
length $l$ of the minimum path between two points separated an
Euclidean distance $R$ scales like $l\propto R^{D_s}$ where $D_s$ is
the fractal dimension of the shortest path. We assume that the
displacement pattern of the frontal region for length less than the
correlation length (in our case $w_s$) is statistically equal to IP
patterns in Ref.~\cite{Sahimi98}.  Therefore, the length of a strand in the
frontal region is proportional to $\Delta h^{D_s}$ when $\Delta h$
is less than $w_s$. If we assume that on the average every tube in the
lattice has same mobility ($k_{ij}/\mu_{ij}$), this causes the fluid
pressure within a single strand to drop like $\Delta h^{D_s}$ as long
as the strand does not split. When the strand splits volume conservation
causes the volume fluxes through the new strands to be less than the 
flux in the strand before it splits. Hence, following a path where
strands split will cause the pressure to drop as $\Delta
h^\kappa$ where $\kappa\le D_s$. From \fig{strands}, we note that at
high $C_a$ the lengths of individual strands in the frontal region
approach the minimum length due to the tubes. Therefore, in this limit
finite size effects are expected to cause $D_s\rightarrow 1$.

From the above arguments we conclude that the pressure drop $\Delta
P_{nw}$, in the nonwetting phase of the frontal region (that is the
strands) should scale as $\Delta P_{nw}\propto\Delta h^\kappa$ where
$\kappa\le D_s$. In 2D two different values for $D_s$ have been
reported: $D_s=1.22$~\cite{Cieplak96,Porto97} for loopless IP with and
without trapping and growing around a central seed, and
$D_s=1.14$~\cite{Sahimi98} for the single strand connecting the inlet
to the outlet when nonwetting fluid percolates the system.  In 2D the
result of {\em Xu} et al.~\cite{Xu98} would give
$\kappa=t/\nu+D_{\text{E}}-1-\beta/\nu\approx 1.9$ where we have
inserted $t=1.3$, $\nu=4/3$, $\beta=5/36$, and $D_{\text{E}}=2$. This result 
for $\kappa$ is larger than $D_s$ and therefore not compatible
with our arguments.

To confirm numerically that $\kappa\le D_s$, we have calculated the
difference in capillary pressure $\Delta P_c$ between menisci along
the front by using our network model. $\Delta P_c$ as function of $\Delta
h$ is calculated by taking the mean of the capillary pressure
differences between all pairs of menisci along the front, separated a
distance $\Delta h$ in the direction of the displacement. $\Delta P_c$
of a pair of menisci is calculated by taking the capillary pressure of
the meniscus closest to the inlet minus the capillary pressure of the
meniscus closest to the outlet. 

\begin{figure}
\begin{center}
\mbox{\psfig{figure=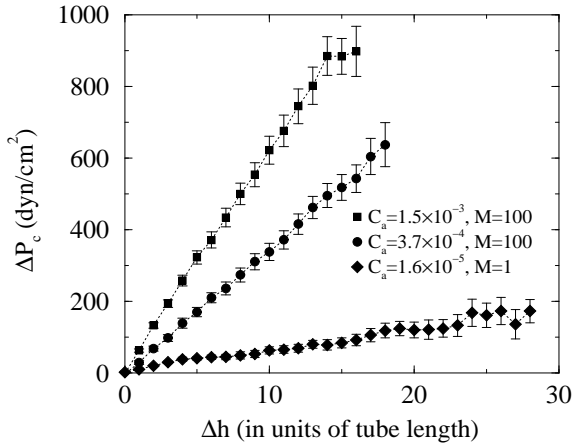,width=7.5cm}}
\caption{$\Delta P_c$ as function of $\Delta h$ for three
different $C_a$'s with $M=100$ and $1$ on lattices of $25\times 35$ and
$40\times 60$ nodes, respectively.}
\label{fig:pc}
\end{center}
\end{figure}
Figure~\ref{fig:pc} shows $\Delta P_c$ as function of $\Delta h$ for
simulations performed at high, intermediate and low $C_a$, with $M=1$
or $100$. The simulations with $M=100$ were performed on a $25\times
35$ nodes lattice with $\mu_{nw}=10$ P, $\mu_w=0.10$ P, and
$\gamma=30$ dyn/cm. The disorder was introduced by choosing the tube
radii at random in the interval $0.05d\le r_{ij}\le d$. The tube
length was $d=0.1$ cm. The simulations with $M=1$ were performed on a
distorted lattice of $40\times 60$ nodes where $0.02\ \mbox{cm}\le
d_{ij}\le 0.18$ cm and $r_{ij}=d_{ij}/2\alpha$ with
$\alpha=1.25$. Here $\mu_{nw}=\mu_w=0.5$ P. We did 10--30 simulations
at each $C_a$ to obtain reliable average quantities. From the plot we
observe that $\Delta P_c$ increases roughly linearly as function of
$\Delta h$. The simulations also show that $\Delta P_c\simeq\Delta
P_{nw}-\Delta P_w$ and that $\Delta P_w$ depends linearly on $\Delta
h$ due to the compact wetting phase there (see \fig{model}). Hence our
simulations give that $\Delta P_{nw}\sim \Delta P_c \propto \Delta
h^\kappa$ where $\kappa\simeq 1.0$. This supports the arguments giving
$\kappa\le D_s$, and we conclude that earlier proposed
theories~\cite{Wilk86,Len89,Blunt92,Xu98} which do not consider the
evidence that nonwetting fluid flows in strands, are incompatible with
drainage when strands are important.

To save computation time and thereby be able to study $\Delta P_c$ on
larger lattices in the low $C_a$ regime, we have generated bond
invasion percolation (IP) patterns with trapping on lattices of
$200\times 300$ nodes. The bonds in the IP lattice correspond to the
tubes in the network model. The occupation of bonds started at the
bottom row, and the next bond to be occupied was always the bond with
the lowest random number $f$ from the set of empty bonds along the
invasion front. We applied a small gradient in the random numbers of the
bonds to stabilize the front~\cite{Wilk86,Birov91}. 

When the IP patterns became well developed with trapped (wetting)
clusters of sizes between the bond length and the front width, the
tubes in our network model were filled with nonwetting and wetting
fluid according to occupied and empty bonds in the IP
lattice. Furthermore, the radii $r$ of the tubes were mapped properly
to the random numbers $f$ of the bonds by first removing the gradient
in $f$ and then assigning $r=1-f$~\cite{Aker00-PRE}. The last
transformation is necessary because in the IP algorithm the next bond
to be invaded is the one with the lowest random number, opposite to
the network model, where the widest tube will be invaded first.

Having initialized the tube network, the network model was started and
the simulations were run a limited number of time steps while
recording $\Delta P_c$. The number of time steps where chosen
sufficiently large to let the menisci along the front adjust according
to the viscous pressure set up by the injection rate. By this method
we save the computation time that would have been required if the
displacement patterns should have been generated by the network model
instead of the much faster IP algorithm. However, to make this method
self-consistent we have to assume that the IP patterns are
statistically equal to the corresponding structures that would have
been generated by the network model.

\begin{figure}
\begin{center}
\mbox{\psfig{figure=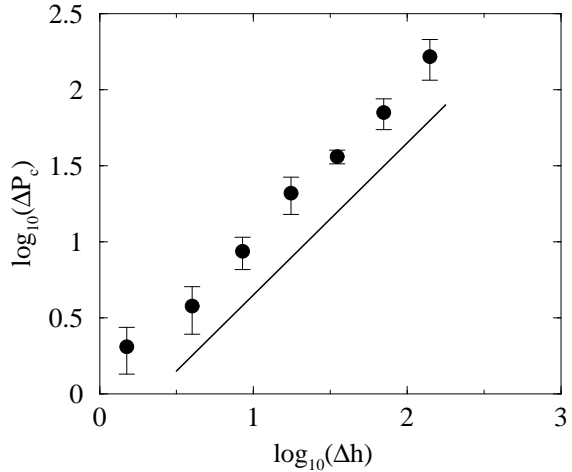,width=7.5cm}}
\caption{The logarithm of $\Delta P_c$ as function of the logarithm of
$\Delta h$ for simulations initiated on IP patterns at
$C_a=\protect\exff{9.5}{-5}$ and $M=100$. The slope of the solid line
is 1.0.}
\label{fig:pcd-ip}
\end{center}
\end{figure}

Totally, we generated four IP patterns with different sets of random
numbers and every pattern was loaded into the network model.  The
result of the calculated $\Delta P_c$ versus $\Delta h$ is shown in
\fig{pcd-ip} for $C_a=\exf{9.5}{-5}$ and $M=100$.  If we assume a
power law $\Delta P_c\propto \Delta h^\kappa$, we find $\kappa =
1.0\pm 0.1$. The slope of the straight line in \fig{pcd-ip} is 1.0.
We have also calculated $\Delta P_c$ for $C_a=\exf{2}{-6}$ with $M=1$
and $M=100$ by using one of the generated IP patterns. The result of
those simulations is consistent with \fig{pcd-ip} which indeed show the
similar behavior as the results in \fig{pc}.

An important issue, arising at low $C_a$, is the effect of bursts
on the capillary pressure. A burst occurs when a meniscus along the
front becomes unstable and nonwetting fluid abruptly covers new
tubes~\cite{Maloy96}. The tube where the burst initiates will during
the burst, experiences a much higher fluid transport relative to
tubes far away. Describing the pressure behavior between the tube
of the burst and the rest of the front is nontrivial. However,
simulations show that even during bursts, we find that $\Delta
P_c$ increases linearly with $\Delta h$.

The evidence that $\kappa\simeq 1.0$ may influence the exponent
$\alpha$ in $w_s\propto {C_a}^{-\alpha}$.  Assuming Darcy flow where
the pressure drop depends linearly on the injection rate, we
conjecture that $\Delta\widehat{P}_c\propto C_a\Delta
h^{\kappa}$. Here $\Delta\widehat{P}_c$ denotes the capillary pressure
difference over a height $\Delta h$ when the front is stationary. That
means, $\Delta\widehat{P}_c$ excludes situations where nonwetting
fluid rapidly invades new tubes due to local instabilities
(i.e. bursts).  The above conjecture is supported by simulations
showing that in the low $C_a$ regime $\Delta\widehat{P}_c\propto
C_a\Delta h^{\kappa}$ where $\kappa\simeq 1.0$.  Note, that
$\Delta\widehat{P}_c\not\simeq \Delta P_c$ in \fig{pc} since the
latter includes both stable situations and bursts.

At sufficiently low $C_a$ where only the strength of the capillary
pressure decides which tube should be invaded or not, we may map the
displacement process to percolation giving $\Delta\widehat{P}_c\propto
f-f_c\propto\xi^{-1/\nu}$~\cite{Sap88,Wilk86,Birov91}. Here $f$ is the
random numbers in the percolation lattice, $f_c$ is the critical
percolation threshold, and $\xi\propto w_s$ is the correlation
length. Combining the above relations for
$\Delta\widehat{P}_c$ gives $w_s\propto {C_a}^{-\alpha}$ where
$\alpha=\nu/(1+\nu\kappa)$. In 2D $\nu=4/3$ and by inserting
$\kappa=1.0$ we obtain $\alpha\approx 0.57$. Note that this is
different to results suggested in Refs.~\cite{Wilk86,Blunt92,Xu98} giving
$\alpha\approx 0.37$--$0.38$ in 2D.

At high $C_a$ the nonwetting fluid is found to invade simultaneously
everywhere along the front, and consequently the front never reaches a
stationary state~\cite{Aker00-PRE}. In this limit simulations show a
nonlinear dependency between $\Delta\widehat{P}_c$ and
$C_a$. Therefore, in the high $C_a$ regime it is not clear if the
above mapping to percolation is valid, and we expect another type of
relation between $w_s$ and $C_a$.

\subsection{COMPARISON WITH EXPERIMENTS}

\begin{figure}
\begin{center}
\mbox{\psfig{figure=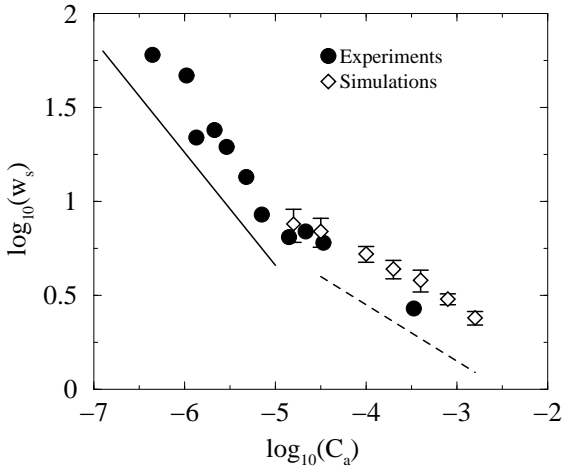,width=7.5cm}}
\caption{$\log_{10}(w_s)$ as function of $\log_{10}(C_a)$ for
experiments from~\protect\cite{Frette97} and simulations on the
lattice of $40\times 60$ nodes. In both experiments and simulations
$M=1$. The slope of the solid and dashed line is -0.6 and -0.3,
respectively.}
\label{fig:w-Ca}
\end{center}
\end{figure}
{\em Frette} et al.~\cite{Frette97} have performed two-phase drainage
experiments in a 2D porous medium with viscosity matched fluids
($M=1$).  They reported on the stabilization of the front and measured
the saturated front width $w_s$, as function of $C_a$.  Their best
estimate on the exponent in $w_s\propto {C_a}^{-\alpha}$ was
$\alpha=0.6\pm0.2$. This is consistent with the above conjecture
($\alpha=0.57$), however, corresponding simulations on $40\times 60$
nodes lattices give $\alpha=0.3\pm0.1$. Figure~\ref{fig:w-Ca} contains
the data of {\em Frette} et al. (filled circles) and the result of our
simulations (open diamonds).  The simulations are performed at
$C_a\ge\exf{1.0}{-5}$ while most of the experiments where done at
$C_a\le\exf{1.0}{-5}$. Since the range of the two does not overlap it
is difficult to compare the result of the simulations with those of
the experiments.  However, the change in $\alpha$ from 0.6 to 0.3,
might be consistent with a crossover to another behavior at high $C_a$
according to the discussion in \secn{loopless}.  We also note, that in 
the simulations at $C_a\simeq\exf{1.0}{-5}$, the front width
approaches the maximum width due to the system size, making it
difficult to observe any possible $\alpha\approx 0.57$ regime at low
$C_a$. We emphasize that more simulations on larger systems and at
lower $C_a$ are needed before any conclusion on $\alpha$ can be drawn.

\subsection{DISCUSSION\label{sec:loops}}

The evidence that the nonwetting fluid displaces the wetting fluid in
a set of loopless strands opens new questions about the displacement
process. Returning to \fig{strands} it is striking to observe the
different patterns of strands at high and low $C_a$. At low $C_a$ few
strands are supplying the frontal region with nonwetting fluid, and
the strands split many times before the whole front is covered.  At
high $C_a$ the horizontal distance between each strand in the static
structure is much shorter, and only a few splits are required to
cover the front. We conjecture that the average horizontal distance
between the fluid supplying strands depends on the front
width. However, further investigation of the displacement patterns is
required before any conclusions can be drawn.

So far the arguments in \secn{loopless} only consider displacements
where the nonwetting strands contain no loops. A very interesting
question that has to be answered is: What happens to $\kappa$ when
different strands in the front connect to generate
loops. In ordinary bond or site percolation
loops generally occur. Loops are also observed in
experiments corresponding to those of {\em Frette} et
al.~\cite{Frette97}. In the experiments it is more difficult to trap
wetting fluid due to the more complex topology of pores and throats
(see \fig{burst}). Consequently, loops will more easily generate there, than
in the case of a regular square lattice. Loops might also be created
when neighboring menisci along the front overlap and coalesce
depending on the wetting properties of the nonwetting
fluid~\cite{Cieplak88,Cieplak90}.

As a first approximation we conjecture that creation of loops will not
cause $\kappa$ to change significantly. Note that in the front the
different nonwetting strands connecting to each other to create
loops, must at some later time split. Otherwise successive connections
will cause the different strands to coalesce into one single strand of
nonwetting fluid. Moreover, after the front width has saturated, the
number of places where different strands connect must on average be
balanced by the number of places where strands split.  Therefore, we
believe that the influence on $\kappa$ due to connections (i.e. loops)
will be compensated by the splits and the overall behavior of $\kappa$
will remain the same.  We emphasize that further simulations and
experiments are required to investigate the effect of loops on
$\kappa$. 

According to the discussion in \secn{loops}, the evidence that the
displacement patterns consist of loopless strands may easily be
generalized to 3D. Therefore our arguments giving $\kappa\le D_s$,
might be valid in 3D as well.  Note also that in 3D it is less
probable that different strands meet.  Hence, even if they were
supposed to connect to create loops, the number of created loops are
expected to be few. In 3D the fractal dimension of the shortest path
for loopless IP is $D_s=1.42$~\cite{Cieplak96}.

\section{CONCLUSION\label{sec:further}}

We conclude that our 2D network model properly simulates the temporal
evolution of the pressure in the fluids during drainage. We have seen
that capillary forces situated around isolated and trapped regions of
wetting fluid, contribute to the total pressure across the lattice, as
well as the capillary fluctuations due invasion of nonwetting fluid
along the front.  

We have found that the model reproduces the typical burst dynamics at
low injection rates and we have investigated the statistics of the
bursts by calculating the hierarchical valley size distribution
$N_{\rm all}(\chi)$. We conclude that $N_{\rm all}(\chi)$ follows a
power law, $N_{\rm all}(\chi)\propto \chi^{-\tau_{\rm all}}$ where
$\tau_{\rm all}=1.9\pm 0.1$ is independent of the injection rate and
viscosity ratio. Similar result is obtained from experiments. At low
injection rates the result is consistent with the prediction in
Ref.~\cite{Maslov95} ($\tau_{\rm all}=2$), which was deduced for a
broad spectrum of different self-organized critical models including
IP. The evidence that $\tau_{\rm all}$ is independent of the injection
rate, may indicate that the displacement process belongs to the same
super universality class as the self-organized critical models in
Ref.~\cite{Maslov95}, even where there is no clear mapping between the
displacement process and IP.

We have also simulated the behavior of the capillary pressure along
the front. Simulations show that the capillary pressure difference
$\Delta P_c$ between two points along the front varies almost linearly
as function of distance $\Delta h$ in the direction of the
displacement. The numerical result supports arguments based on the
observation that nonwetting fluid flows in separate strands where
wetting fluid is displaced. From the arguments we find that $\Delta
P_{c}\propto \Delta h^\kappa$ where $\kappa\le D_s$. Here $D_s$ denotes
the fractal dimension of the nonwetting strands.

Several attempts have been made to de\-scribe the stabilization
mechanisms in drain\-age due to viscous forces, however, none of them
consider the evidence that nonwetting fluid displaces wetting fluid
through strands. Therefore, we conclude that earlier suggested
theories fail to describe the stabilization of the invasion front when
strands dominate the displacements.

\section*{ACKNOWLEDGMENTS}

E.~Aker and A.~Hansen wish to thank the Niels Bohr Institute and the Nordic
Institute for Theoretical Physics for their hospitality. The work has received
support from the Norwegian Research Council through a ``SUP'' program 
and from the Niels Bohr Institute.

\end{document}